\documentclass{article}
\usepackage{hyperref}
\usepackage[utf8]{inputenc}
\usepackage{amsmath}
\usepackage[left=1.50cm,
right=1.50cm,
top=2.00cm,
bottom=2.00cm
]{geometry}
\usepackage{graphicx}
\usepackage{float}
\usepackage[backend=biber, style=numeric, citestyle=numeric, sorting=none]{biblatex}
\AtEveryBibitem{\clearfield{note}}
\usepackage{bbold}
\usepackage{caption}
\usepackage{subcaption}
\usepackage{appendix}
\usepackage{braket}
\usepackage{comment}
\usepackage{multicol}
\usepackage{multirow}
\usepackage{comment}
\usepackage{xcolor}
\usepackage{authblk}
\newtheorem{theorem}{Theorem}

\usepackage{amssymb}
\usepackage{enumitem}
\usepackage{array}
\usepackage{authblk}
\usepackage{caption}
\usepackage{titlesec}
\titleformat{\section}
  {\normalfont\large\bfseries}{\thesection}{1em}{}

\titleformat{\subsection}
  {\normalfont\normalsize\bfseries}{\thesubsection}{1em}{}

\titleformat{\subsubsection}
  {\normalfont\small\bfseries}{\thesubsubsection}{1em}{}

\begin{document}

\title{\vspace{-1cm}\large \textbf{Impossibility of universal work extraction from coherence: \\ Reconciling axiomatic and resource-theory approaches}}

\author[1]{\small Samuel Plesnik \thanks{saumpio@gmail.com}}
\author[1,2]{\small Maria Violaris \thanks{maria@violaris.com}}

\affil[1]{\small \textit{Mathematical Institute, University of Oxford, Woodstock Road, Oxford OX2 6GG, United Kingdom}}
\affil[2]{\small \textit{Clarendon Laboratory, University of Oxford, Parks Road, Oxford OX1 3PU, United Kingdom}}

\date{\vspace{-1cm}\small \today}

\normalsize

\maketitle

\renewcommand{\abstractname}{}
\begin{abstract}
\vspace{-1cm} We compare how the impossibility of a universal work extractor from coherence arises from different approaches to quantum thermodynamics: an explicit protocol accounting for all relevant quantum resources, and axiomatic, information-theoretic constraints imposed by constructor theory. We first explain how the impossibility of a universal work extractor from coherence is directly implied by a recently proposed constructor-theoretic theorem based on distinguishability, which is scale- and dynamics- independent. Then we give an explicit demonstration of this result within quantum theory, by proving the impossibility of generalising a proposed quantum protocol for deterministically extracting work from coherence. We demonstrate a new connection between the impossibility of universal work extractors and constructor-based irreversibility, which was recently shown using the quantum homogenizer. Finally we discuss additional avenues for applying the constructor-theoretic formulation of work extraction to quantum thermodynamics, including the irreversibility of quantum computation and thermodynamics of multiple conserved quantities. 
\end{abstract}

\begin{multicols}{2}
    \setlength{\parskip}{0pt}
    \setlength{\parindent}{15pt}

\section{Introduction}

The laws of thermodynamics have a powerful generality, dating back to early expressions of the second law by Clausius \cite{clausius1854veranderte} and Kelvin \cite{thomson1852ii}.  Later, seminal works by Carathéodory \cite{caratheodory1907variabilitatsbereich}, Giles \cite{giles} and  Lieb and Yngvason \cite{lieb1999physics} formalised an axiomatic approach to classical thermodynamics. In parallel, the formulation of statistical mechanics, and recent applications of thermodynamics to microscopic, non-equilibrium and quantum systems, has led to a better understanding of thermodynamic laws in a range of physical contexts \cite{vinjanampathy2016quantum}.

In this work we consider two main approaches to quantum thermodynamics. One is based on the thermodynamics of individual quantum systems, with foundations in resource theories. By explicitly accounting for all the resources involved in manipulating a given quantum system, including back-reaction on ancillary quantum systems, it is possible to account for all the resources of a given kind required for a particular task \cite{brandao2013resource, ng2019resource}. The other approach is information-theoretic in nature, based on the recently proposed constructor theory \cite{deutsch2013constructor, deutsch2015constructor}. The theory aims to express all fundamental laws of physics in terms of principles about possible and impossible tasks, which are ultimately non-probabilistic, scale-independent and dynamics-independent. The constructor-theoretic principles about information and thermodynamics then impose testable constraints on theories with specific dynamical laws, such as quantum theory. 

In particular, we investigate a case-study to show how the resource theory and constructor theory approaches reach intersecting conclusions: the impossibility of a universal work-extractor from coherence. By analysing a proposed protocol for extracting work from coherence \cite{korzekwa2016extraction}, we demonstrate within quantum mechanics that there cannot be a generalisation of the protocol to achieve a universal work extractor. This forbids deterministic work extraction of different amounts of work from any pair of non-orthogonal input states, including coherent and incoherent states of the same energy. We explain how this constraint on work extraction directly follows from a recent dynamics-independent theorem derived in constructor theory \cite{marletto2022information}. Our results show the information-theoretic origins of the impossibility of a universal work extractor from coherence, for quantum theory and its potential successors. During our analysis, we explain how the phenomenon of ``work-locking", whereby the additional free energy in a coherent state is not accessible, emerges as a consequence of constructor-theoretic distinguishability principles being applied to quantum theory. 

Furthermore, we find a new connection between the impossibility of universal work extractors and the recent constructor-theoretic analysis of the quantum homogenizer. Past work has used the quantum homogenizer to demonstrate constructor-based irreversibility, which is a novel, exact form of irreversibility consistent with unitary quantum dynamics \cite{violaris2022irreversibility, marletto2022emergence}. Here we show that the constructor-theoretic impossibility of a universal work extractor in quantum theory relies on the constructor-theoretic impossibility of transforming a qubit from a mixed state to a pure state, which has been shown to hold when the task is implemented by the quantum homogenizer.

We also comment on additional promising avenues for connecting constructor-theoretic results with other approaches in quantum thermodynamics. This could help generalise existing results in quantum thermodynamics to potential successor theories of quantum mechanics, and elucidate the principles from which these phenomena originate. We conjecture an information-theoretic foundation for the additional entropy dissipation that makes quantum computation fundamentally irreversible \cite{bedingham2016thermodynamic}, and suggest how constructor-theoretic results could be extended to multiple conserved quantities. This could lead to new connections between conservation laws and thermodynamic-type laws. 

\section{Comparing resource- and constructor-theoretic approaches} \label{resource_comparison}

We begin with some brief comments on the aims and formulation of resource theories and constructor theory, as applied to thermodynamics, with a more detailed discussion in Appendix \ref{resource_comparison_appendix}.

Resource theories formalize scientific fields by identifying ``free" (unlimited) and ``resourceful" (finite) states and transformations, useful in contexts where resources are distinguished by their availability \cite{coecke2016mathematical}. These theories have significantly contributed to quantum thermodynamics, enhancing understanding of nanoscale thermodynamics and the limits of work extraction \cite{brandao2013resource, horodecki2013fundamental,ng2015limits,brandao2015second, aaberg2013truly,skrzypczyk2014work,renes2014work}.

Constructor theory offers an axiomatic approach to thermodynamics, based on which tasks are possible or impossible on systems whose dynamics may be partially unknown \cite{deutsch2013constructor,deutsch2015constructor,marletto2016constructor}. The central conjecture is that fundamental laws of physics can be expressed as \textit{possible} and \textit{impossible} transformations, offering insights into irreversibility, the nature of information and non-classicality \cite{marletto2020witnessing, di2022temporal, marletto2016constructor, violaris2022irreversibility, marletto2022emergence}.

While resource theories take existing dynamical laws of physics as a starting point and derive constraints on what can be done under those laws, constructor theory has its own laws of physics, for instance the \textit{interoperability of information} and the \textit{principle of locality}. These laws are conjectured to be more fundamental than dynamical laws.\\

\textbf{State transformations vs possible tasks.} The possibility of a task in constructor theory depends on a task being implemented in a cycle (i.e. with no irreducible changes in the environment), with the machine transforming the system in a cycle called the \textit{constructor} \cite{deutsch2015constructor}. The constructor should be approximately unchanged after causing the desired change in the system, and there is no limit to how well a perfect constructor for the task can be approximated. 

In resource theories, catalysts are central to facilitating transformations. Broadly defined, a catalyst is an entity that facilitates a transformation that could not occur without it, and returns to its initial state at the end of the process. Specific definitions and conditions for catalysts can involve considerations such as its potential to form correlations with the system it interacts with, and the acceptability of small deviations from its initial state (e.g. \cite{rubboli2022fundamental}). A key aspect of this approach is therefore the reusability of catalysts without significant alterations to their state.

Contrastingly, constructor theory involves a wider class of entities that cause transformations, called constructors. The key property of a constructor is its capacity to repeatedly perform a given task, without necessarily returning to its exact initial state or a state close to it. The possibility of a constructor for a task depends on how far a system in some input subspace can be repeatedly transformed to a state in an output subspace. This relates to the indirect way in which constructors emerge in constructor theory: the actual laws of physics, stated as possible and impossible tasks, do not refer to specific properties of the entities that implement them (the constructors). The principles instead constrain the limits of how closely an ideal entity that repeatedly performs a task can be approximated by physical machines. 

Recently, the implications of constructor-theoretic principles of thermodynamics have been explored within the specific dynamics of quantum theory \cite{violaris2022irreversibility, marletto2022emergence}. In this setting, existing results and tools from quantum resource theories may have interesting connections with the constructor-theoretic possibility of thermodynamic tasks, after accounting for the subtleties regarding the above mentioned characterisation of constructor-theoretic possibility and the generalisation of catalysts to constructors. \\

\textbf{Connection to Lieb and Yvangson's axiomatic thermodynamics.} Lieb and Yvnangson's axiomatic approach is formulated around the property of ``adiabatic accessibility": the key relation between two equilibrium states is whether or not one is adiabatically accessible from the other, where adiabatic accessibility is defined operationally using a set of axioms \cite{lieb1999physics}. A state is adiabatically accessible from another state if it is possible for the transformation to be implemented by some auxiliary system that returns to its original state, with the only side-effect being the lifting or lowering of a weight in a gravitational field. 

The \textit{resource theory of noisy operations} coincides with the axioms of Lieb and Yvangson applied to quantum theory, leading to \textit{majorization} being the mathematical criterion for computing the adiabatic accessibility of one quantum state from another \cite{weilenmann2016axiomatic}. Meanwhile, the constructor theory of thermodynamics extends the adiabatic accessibility condition to one of \textit{adiabatic possibility}, defined in terms of tasks rather than states, such that it is dynamics-independent and scale-independent \cite{marletto2016constructor, marletto2022emergence}. Table \ref{table:thermo_relations} gives a summary of the approaches to ordering states according to the 2$^{\textrm{nd}}$ law in thermodynamics, and their associated domain of applicability. \\

\begin{table*}[ht]
\centering
\begin{tabular}{|>{\raggedright\arraybackslash}p{0.2\linewidth}|l|>{\raggedright\arraybackslash}p{0.5\linewidth}|}
\hline
\textbf{Approach} & \textbf{Ordering relation} & \textbf{Domain of applicability} \\ \hline
Classical axiomatic thermodynamics & Adiabatic accessibility & Defined in terms of behaviour of a weight in a gravitational field. Scale-dependent.\\ \hline
Constructor theory of thermodynamics & Adiabatic possibility & Defined entirely using possible and impossible tasks, can be applied to settings of systems instantiating classical and/or quantum and/or more general forms of information. Scale- and dynamics-independent \\ \hline
Resource theory of noisy quantum operations & Majorization & Defined for quantum states, with a ``weight" understood as a work storage medium independent of a specific model. Scale- and dynamics-dependent.\\ \hline
\end{tabular}
\caption{Comparison of how states are ordered in different approaches to thermodynamics}
\label{table:thermo_relations}
\end{table*}

\section{Constructor-theoretic impossibility of work from coherence}

We will briefly review the constructor theory background required to understand the recent theorem connecting work extraction and distinguishability, with full details of the theorem available in \cite{marletto2022information}, and full details of the constructor theory of information in \cite{deutsch2015constructor}. 

In constructor theory, it is conjectured that all fundamental laws of physics can be expressed as statements of what tasks are possible and impossible, and why. Individual systems are called \textit{substrates} and are characterised by their \textit{attributes}, which are sets of all states with a given property. These properties can be changed by physical transformations. A \textit{task} is a transformation of a substrate from an input attribute to an output attribute. A \textit{variable} is a set of disjoint attributes, i.e. a set of distinct, independent properties that a system can have. 

For a system $\mathbf{S_1}$ with attribute $\mathbf{a}$ and $\mathbf{S_2}$ with attribute $\mathbf{b}$, the combined system $\mathbf{S_1} \oplus \mathbf{S_2}$ has attribute $\mathbf{(a, b)}$. An important principle is the \textit{principle of locality}, which states that if a transformation only operates on substrate $\mathbf{S_1}$, then only attribute $\mathbf{a}$ can change \cite{deutsch2015constructor}. 

A \textit{constructor} for a task $\mathbf{T}$ is a system that, when presented with a substrate with one of the input attributes of $\mathbf{T}$, delivers it in one of the output attribute states of  $\mathbf{T}$ with arbitrarily small error, and retains the capacity to repeat this process. A model for a constructor within quantum theory has been proposed in \cite{violaris2022irreversibility}. The constructor is defined as being the limit set, $\mathbf{C}$, of a sequence of sets of quantum systems that implement $\mathbf{T}$ on the substrate. 

A task is \textit{impossible} ($\mathbf{T}^{\times}$) if the laws of physics limit the accuracy of a constructor for that task. Otherwise, if there is no limit short of perfection for a constructor to exist for a task, the task is \textit{possible} ($\mathbf{T}^{\checkmark}$). For a possible task, a perfect constructor can be approximated with arbitrarily good precision by a sequence of ever-improving approximate constructors. 

Constructor theory has been used to express laws of information and thermodynamics in a scale-independent and dynamics-independent way. \textit{Information media} are defined in terms of what tasks are possible and impossible on those media, and \textit{work media} are defined as a special case of information media \cite{marletto2016constructor}. In particular, a novel scale and dynamics-independent definition of thermodynamic work and work extraction was recently proposed \cite{marletto2022information}.  A \textit{deterministic work extractor} can transform the attributes of a substrate to those of a \textit{work variable}, where a work variable is the set of distinct attributes of a work medium. 

Let's consider the task of work extraction of a quantum system as an example. In a quantum system, the work extraction task can be implemented by coupling a system $\rho_S$ (which is the substrate $\mathbf{S}$) to an ancillary battery $\rho_B$ (which is the work medium $\mathbf{Q}$). Then an energy-conserving unitary transformation $U$ can transfer energy between the system and the battery, such that the final state of the battery is $\rho_B^\prime =  \operatorname{Tr}_S U (\rho_S \otimes \rho_B) U^\dagger$. The work medium $\mathbf{Q}$ could be e.g. an atom with multiple energy levels that get excited or de-excited because of the interaction with a substrate $\mathbf{S}$, where $\mathbf{S}$ is e.g. a magnetic field or another atom. If both media are atoms, and both their variables are sets of energy eigenstates of the atom, then the deterministic work extractor maps each energy eigenstates of the system $\rho_S$ to a unique energy eigenstate of the battery $\rho_B$.

In quantum theory, single-shot distinguishability of states is possible if they are orthogonal. Using constructor theory, the same kind of single-shot distinguishability can be defined without referring to properties of dynamics specific to quantum theory, such as orthogonality (see Appendix \ref{distinguishability}). Hence the constructor theoretic definition of distinguishability generalises the distinguishability of orthogonal states in quantum theory. In \cite{marletto2022information}, the constructor-theoretic formulation of distinguishability is used to prove that work variables are distinguishable, theorem \ref{theorem:distinguish}:

\begin{theorem}
\label{theorem:distinguish}
  A work variable $\mathbf{W}$ is a distinguishable variable.
\end{theorem}

In general, a deterministic work extractor transforms each attribute of the substrate $\mathbf{S}$ to exactly one attribute of the work medium $\mathbf{Q}$. Hence, since $\mathbf{Q}$'s variable is distinguishable, $\mathbf{S}$ must also have a distinguishable variable \cite{deutsch2015constructor}. 

\subsection{Constraints on coherence}

When applied to quantum theory, theorem \ref{theorem:distinguish} means that it is possible to deterministically extract different amounts of work from two quantum states only if they are orthogonal. If two quantum states are non-orthogonal, then it could be possible to deterministically extract the same amount of work from the two states. However, the theorem rules out having a universal deterministic work extractor able to extract different amounts of work from unknown, non-orthogonal states, visualised in Figure \ref{work_coherence}. This theorem has interesting consequences for (in)coherent inputs, since a coherent state and incoherent state of the same energy are in general non-orthogonal, and so not perfectly distinguishable in a single shot. Hence, the theorem forbids a work extractor that can deterministically extract more work from a coherent state than the corresponding incoherent state of the same energy, despite the former having greater free energy. 

Specifically, consider a system with two energy eigenstates, $\ket{0}$ and $\ket{1}$. The system can be in a coherent state, $\ket{\psi} = \alpha \ket{0} + \beta \ket{1}$, with $\rho_{\text{coherent}} = \left( \begin{smallmatrix} |\alpha|^2 & \alpha\beta^* \\ \beta\alpha^* & |\beta|^2 \end{smallmatrix} \right)$ or the system can be in an incoherent state of the same energy, with $\rho_{\text{incoherent}} = \left( \begin{smallmatrix} |\alpha|^2 & 0 \\ 0 & |\beta|^2 \end{smallmatrix} \right)$.\\

\begin{center}
    \includegraphics[width=\linewidth]{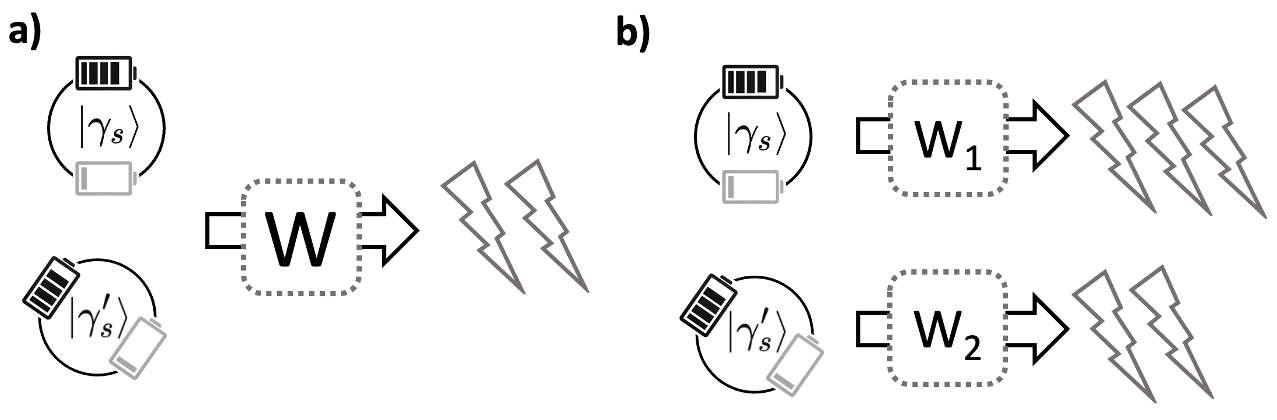}
    \captionof{figure}{Consider deterministic work extraction from non-orthogonal coherent states, $\ket{\gamma_s}$ and $\ket{\gamma'_s}$. a) A given machine could extract the same amount of work from both inputs. b) Extracting different amounts of work requires special-purpose machines.}
    \label{work_coherence}
\end{center}

Let's denote the task of extracting a quantity of work $w$ from a quantum state $\rho$ as $\mathbf{T_w}( [\rho, w])$. Assuming the task is possible, let's denote a constructor capable of causing the task as $\mathbf{C_w}([\rho, w])$. The task of deterministically extracting amounts of work $w_1$, $w_2$ from a pair of states $\rho_1$, $\rho_2$ respectively is $\mathbf{T_w}([\rho_1, w_1], [\rho_2, w_2])$. A consequence of the constructor-theoretic theorem is that $\mathbf{T_w}$ is possible ($\mathbf{T_w^{\checkmark}}$) if and only if $\rho_1$ and $\rho_2$ are perfectly distinguishable with a single shot measurement, i.e. are orthogonal quantum states. These orthogonal states could be energy eigenstates, or a pair of orthogonal states coherent in the energy basis. The theorem therefore does not permit a constructor $\mathbf{C_w}([\rho_1, w_1], [\rho_2, w_2]))$ for extracting different amounts of work from non-orthogonal states. This includes both arbitrary pairs of non-orthogonal coherent states, and pairs of coherent  and incoherent states of the same energy. Consequently, there cannot be a universal work extractor capable of distinguishing between and extracting different amounts of work from such states, highlighting a fundamental limitation in the physics of work extraction from quantum coherence. 

We demonstrate this explicitly using the coherent work extraction protocol proposed in \cite{korzekwa2016extraction}. We focus on the pre-processing step, where knowledge of the input state is important, and show how this fails to generalise for sets of non-orthogonal inputs. This demonstrates the physical significance of prior knowledge of input states for constraining extractable work, which has been explored in various quantum settings (e.g. \cite{vsafranek2023work}).

\section{Impossibility of universal quantum work from coherence}

\subsection{Resource theory background}

The resource theoretic approach is used to analyse work extraction from coherent states in \cite{skrzypczyk2014work}, where it is shown that the average work extracted from a single copy of a mixture of energy eigenstates can be equal to the free energy of the state. For coherent states, if no external sources of coherence are used, collective actions on multiple copies of the state are needed to extract work equal to the free energy. However, for an individual coherent state, no more work can be extracted than that for its incoherent counterpart of the same energy, despite the former having more free energy. This phenomenon is known as \textit{work-locking}.

It is then shown in \cite{aaberg2014catalytic} that a coherent resource can be repeatedly used as a catalyst for tasks involving a change in coherence of a system. Extending this set up to the problem of extracting work from coherent states, in \cite{korzekwa2016extraction} it is shown that thermal machines can extract single-shot work from coherent states with arbitrarily small failure probability, though the work extracted is non-optimal when using a bounded coherent catalyst.  Hence, adding a coherent resource enables one to bypass the constraints of work-locking. We review the assumptions and setting of work-locking and the protocol in \cite{korzekwa2016extraction} in Appendix \ref{work_protocol}. Here we summarise the steps for extracting work from coherence.

Work can be extracted from the coherence in the system by following a three-step protocol: 1) pre-processing the system into a state that allows optimal work extraction; 2) work extraction; and 3) resetting the ancilla, to enable a cyclic process, as summarised in Figure \ref{fig:korzewaPaper} a). The protocol involves a system state $\rho_S$, a battery state $\rho_B$, and a reference state $\rho_R$. The pre-processing step applies a desired channel to $\rho_S$ using $\rho_R$ as an ancilla; the work extraction step involves extracting work from $\rho_S$ into $\rho_B$; and finally to ensure repeatability, the reference is ``repumped" to be close to original state, via a joint operation on $\rho_S$, $\rho_B$ and $\rho_R$. 

The pre-processing step requires knowledge of the system. Specifically, the operation involves rotating the state $\ket{\psi}$,
\begin{equation}
\label{eq:CoherenceInputstate}
    \ket{\psi} = \sqrt{1-p}\ket{0} + \sqrt{p}e^{-i\varphi}\ket{1},
\end{equation}
parameterised by $p$ and $\varphi$, into a coherent Gibbs state $\ket{\gamma}$ with thermal distribution $(1-r,r)$,
\begin{equation}
\label{eq:coherentGibbsState}
    \ket{\gamma} = \sqrt{1-r}\ket{0} + \sqrt{r}\ket{1}.
\end{equation}
This rotation is defined by the following quantum channel:
\begin{equation}
    \ket{\psi} \rightarrow \ket{\gamma}: \qquad \qquad Tr_R \left[ V \left( \ket{\psi}\bra{\psi} \otimes \rho_R \right) V^\dagger \right] = \ket{\gamma}\bra{\gamma},
\end{equation}
where $V(U)$ is the energy-preserving unitary. $V(U)$ is used to approximately induce the transformation $U$ on $\ket{\psi}$, 
\begin{align}
    V(U) &= \ket{0}\bra{0} \otimes \ket{0}\bra{0} + \sum_{l=1}^\infty V_l(U) \\
    V_l(U) &= \sum_{n,m = 0}^1  \bra{n} U \ket{m} \ket{n}\bra{m} \otimes \ket{l-n}\bra{l-m}.
\end{align}

$U$ is chosen to rotate state $\ket{\psi}$ into $\ket{1}$,
\begin{equation} \label{eq: U}
    U = \begin{pmatrix} \sqrt{p} & -\sqrt{1-p} \\ \sqrt{1-p} & \sqrt{p} \end{pmatrix}.
\end{equation}

Once the pre-processing stage is complete, the system and reference states are denoted as $\rho_S^\prime$ and $\rho_R^\prime$, respectively. The subsequent step involves the extraction of work. This is accomplished by applying the transformation $W(\ket{\gamma}\bra{\gamma}\otimes \rho_{battery})W^\dagger$. This transformation aims to extract a desired amount of work, denoted as $w$, with a probability of success of $1-r$. The probability of failing to extract work is $r$.

To ensure the repeatability of the protocol, it is necessary to make adjustments to the reference state $\rho_R^\prime$. This adjustment is made by considering the energy transferred to the battery. A repumping process is performed to correct the reference state, costing one unit of work. The resulting state serves as the corrected reference state, which can be used for subsequent repetitions of the protocol, ensuring that work can be extracted in a cycle. We discuss some caveats regarding the independence of the output systems from these kinds of protocols in Appendix \ref{catalyst_caveats}. 

\subsection{Connections between constructor-theoretic distinguishability and work-locking}

We now consider how work-locking emerges as a natural consequence of the constructor-theoretic theorem \ref{theorem:distinguish}, including limits on its domain of applicability. Work-locking refers to the phenomenon where without access to an external source of coherence, the same work is extractable from both a coherent and incoherent state of the same energy. The additional free energy of the coherent state cannot be converted into work. For the case of deterministic work extraction, a related phenomenon is implied more generally by theorem \ref{theorem:distinguish}: if different amounts of work were deterministically extractable from coherent and incoherent inputs, this would violate the theorem, as the work-extractor would act as a perfect distinguisher of inputs that are not perfectly distinguishable. 

However, the constraint of work-locking can be evaded to extract work that harnesses the free energy of the coherent state, in a regime where an asymptotic number of systems are processed (shown in section IV. b) ``Extracting work arbitrarily close to the free energy difference" in \cite{korzekwa2016extraction}). Theorem \ref{theorem:distinguish} is based on the impossibility of distinguishing states with a single shot, but in the limit of processing an asymptotic number of systems, any quantum states become perfectly distinguishable. This is the principle behind quantum tomography, expressed more generally as the principle of asymptotic distinguishability in \cite{marletto2022information}. Hence, the fact that the full free energy of the coherent state can be extracted when an asymptotic number of system states are input into the work extractor intersects with the domain of applicability of the constructor-theoretic theorem. In the next section, we discuss the more subtle situation whereby systems are processed in a single shot (not in the asymptotic limit), and external catalytic resources are allowed. 

\subsection{Proof protocol is not universal}

The step in the work-from-coherence protocol where knowledge of the system's state is significant is the pre-processing step and application of the rotation $U$ (eq. \ref{eq: U}). The transformation $U$ is required to rotate the state $\ket{\psi}$ to $\ket{1}$, enabling extraction of maximal work \cite{kammerlander2016coherence}. The protocol uses an initial state coherent in the energy eigenbasis, where $\ket{0}$ has lower energy than $\ket{1}$. The ability to rotate the state from which work is to be extracted to the higher energy $\ket{1}$ state is crucial to extract the optimal amount of work. 

\begin{figure*}[ht]
    \centering
    \includegraphics[width=5.9in]{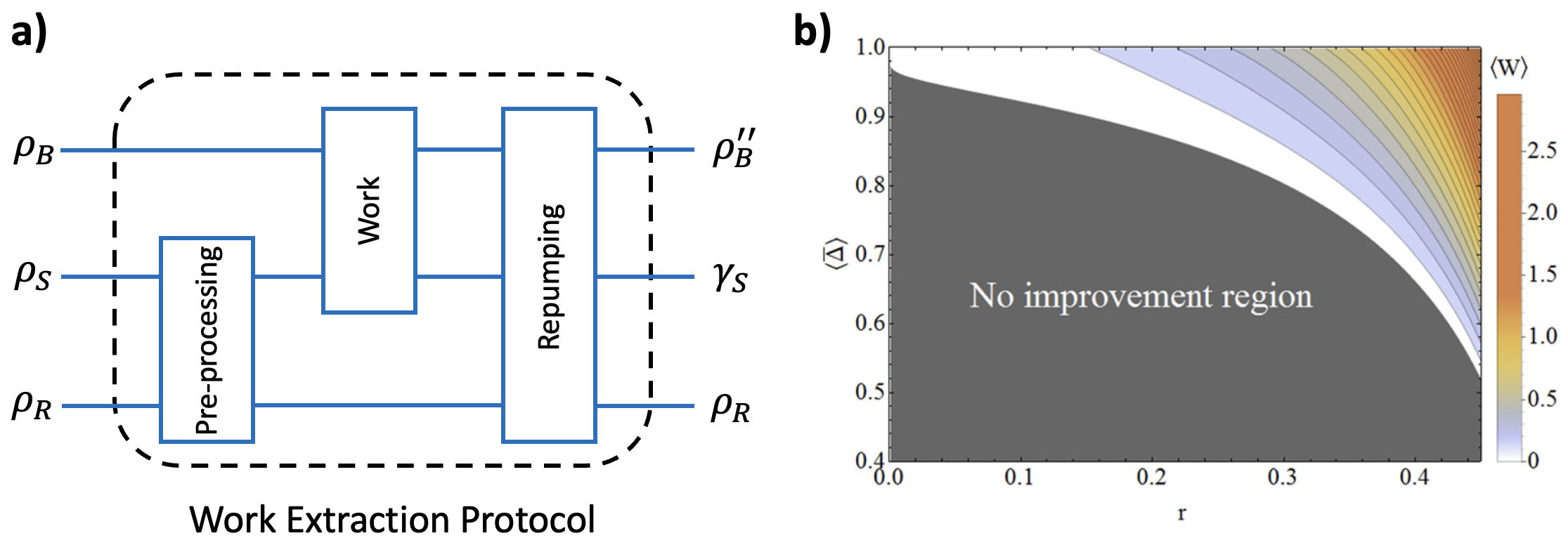}
    \caption{a) Quantum circuit diagram schematic of the work extraction protocol from coherence, where $\rho_B$ and $\rho_B^{\prime \prime}$ are initial states of the battery, $\rho_S$ and $\gamma_S$ are initial coherent state and final Gibbs thermal state and $\rho_R$ is the reference. b) Figure reproduced from \cite{korzekwa2016extraction}, Fig. 4. A phase diagram showing regions where the coherent work extraction protocol provides improvement and is capable of extracting more work from a coherent state than from an incoherent state, where $r$ is the output Gibbs state coherent state parameterisation and $\langle \bar{\Delta} \rangle$ reference quality parameter.}
    \label{fig:korzewaPaper}
\end{figure*}

This is because the quality of the resulting reference state in the protocol affects the amount of work that can be extracted, as the reference quality is defined by the distance of the rotated state from $\ket{1}$. It was shown in \cite{korzekwa2016extraction} that for each coherent Gibbs state parametrised by $r$ (equation \ref{eq:coherentGibbsState}), there exists a minimal quality of the reference state that ensures an advantage in extracting work from a coherent state compared to its incoherent counterpart. In the single-shot case, this advantage is a smaller probability of failure. The failure rate can be made arbitrarily small as the quality of the reference state is improved, tending towards deterministic work extraction. The phase diagram in Figure \ref{fig:korzewaPaper} b) shows the state and reference quality combinations that result in improved work extraction by the protocol. Therefore, mapping the initial state to $\ket{1}$ in the energy eigenbasis is a crucial step in the coherent work extraction protocol, to ensure that the reference state will fall in a region where an improvement in work extracted from the coherent state can be seen. \\

\textbf{Pure, distinct and non-orthogonal input states.} Let's first consider attempting to generalise the protocol so that it can take as input two non-orthogonal, pure coherent states. It is clearly impossible to choose $U$ such that $U\ket{\psi} = \ket{1}$ for two distinct inputs $\ket{\psi}$, as there will be an irreducible distance between their output states. A key property of fidelity $F$ (distance between states) is its invariance under unitary transformations on the state space \cite{jozsa1994fidelity}. For example, the fidelity between $U\ket{0}$ and $U\ket{+}$ is $\frac{1}{\sqrt{2}}$ for any $U$. More generally, for two pure states, consider a unitary transformation $U$ such that $U\ket{\psi_1} = \ket{1}$ and $U\ket{\psi_2}=\ket{1}$ exists for arbitrary two states $\ket{\psi_1}, \ket{\psi_2}$ where $\ket{\psi_1} \neq \ket{\psi_2}$. Then we can write: $1 = \braket{1 | 1} = \bra{\psi_1} U^\dagger U \ket{\psi_2} = \braket{\psi_1 | \psi_2}$. Since $\braket{\psi_1 | \psi_2} = 1$ is true only if $\ket{\psi_1} = \ket{\psi_2}$, this contradicts our initial assumption, so this $U$ cannot exist. \\

\textbf{Orthogonal input states.} A key exception to this is for perfectly distinguishable states. If the two inputs are deterministically, single-shot distinguishable, e.g. $\ket{0}$ and $\ket{1}$, the fidelity of $U\ket{0}$ and $U\ket{1}$ will be 0. However, a measurement can be done in the basis in which the inputs are perfectly distinguishable, and then the appropriate $U$ applied conditionally to the input, depending on its state. \\

\textbf{Allowing an error tolerance.} The impossibility of generalising the protocol is not quite so clear cut, since there is some error tolerance built in: $V(U)$ only needs to approximately implement $U$ on $\ket{\psi}$ to gain an advantage in work extraction from coherence, giving a final system state:

\begin{equation}
    \rho_S^\prime \approx U (\ket{\psi}\bra{\psi})U^\dagger = \ket{1}\bra{1}.
\end{equation}

This means that there is some tolerance for the protocol to work for multiple, non-orthogonal input states. The input states must be close enough that $V(U)$ approximately maps the system to $\ket{1}\bra{1}$, with the accuracy of the approximation lower bounded by the distance from $\ket{1}\bra{1}$ admissible for work to be extracted from coherence. Specifically, the effectiveness of applying $V(U)$ is quantified by a parameter $q$ (see equation 11 in \cite{korzekwa2016extraction}), with $q = 1$ indicating an ideal channel mapping the system to $\ket{1}\bra{1}$. If $U\ket{\psi} = \ket{1}$, then $q$ can be made arbitrarily close to 1.  If $U\ket{\psi} = \epsilon \ket{0} + \sqrt{1-\epsilon}\ket{1}$ for some small $\epsilon$, then the final system state $\rho'_s$ has an $O(\epsilon)$ correction, leading to $q$ having an $O(\epsilon)$ correction. So, the protocol can be applied for multiple non-orthogonal inputs for the constrained set of states where coherent work extraction is possible within an $O(\epsilon)$ approximation.  

By definition, a universal machine must work for any input state. There will always be possible input states that have a large distance between them (e.g. those that are close to orthogonal). Yet, non-orthogonal inputs cannot be perfectly distinguished in a single-shot measurement, meaning that the unitary cannot be made conditional on the input state; it must operate with no specific knowledge of the input state being retrieved. The step of the protocol where $V(U)$ must approximately map any input $\ket{\psi}$ to a state close to $\ket{1}$ will have a fundamentally restricted set of input states, forbidding universality of such a machine. Therefore, this is the specific part of the work extraction protocol that embodies the constraint introduced by the quantum instance of the constructor-theoretic theorem, where $\mathbf{T_w}([\rho_1, w_1], [\rho_2, w_2])$ is possible only if $\rho_1$ and $\rho_2$ are perfectly distinguishable. \\

\textbf{Mixed states as input.} We considered the impossibility of unitarily mapping two arbitrary pure states to a region within the $\ket{1}\bra{1}$ state required for the deterministic work extraction from coherence. However, to deduce the possibility of a deterministic work extractor that can take in both a coherent pure state and an incoherent thermal state of the same energy as input, we must consider the effect of inputting a mixed state to the protocol. In general, two arbitrary, distinct input mixed states $\rho_1$, $\rho_2$ cannot be unitarily transformed to $U\rho_1U^\dagger = U\rho_2U^\dagger = \ket{1}\bra{1}$, again since $1 = F(U\rho_1U^\dagger, U\rho_2U^\dagger) = F(\rho_1, \rho_2) \rightarrow \rho_1 = \rho_2$ , contradicting the assumption of distinct $\rho_1$ and $\rho_2$.

However, there is a more pertinent problem with attempting to use the pre-processing step on the incoherent state corresponding to a coherent state of the same energy: it involves mapping a mixed state to a pure state. This cannot be done unitarily, since it involves an irreducible change in von Neumann entropy. We now consider approximating the mapping of mixed states to pure states using external resources. 

\subsection{Role of catalysis}

We have shown that $V(U)$ cannot directly implement a universal channel for mapping arbitrary input systems to the set of states close to $\ket{1}\bra{1}$. However, there are many quantum channels that become available by the use of catalysts or approximate catalysts, as demonstrated in quantum thermodynamics and quantum resource theories \cite{ng2015limits, lipka2021all}. In particular, various channels are made approximately possible with the introduction of catalysts. 

As noted in section \ref{resource_comparison}, certain aspects of catalysis are generalised by that of a constructor. We will now draw upon recent results that demonstrate the constructor-theoretic impossibility of transforming mixed states to pure states (i.e. impossibility of performing the transformation in a cycle), to show that the coherent work extraction protocol cannot be made universal even when taking advantage of a candidate constructor for the task. 

\section{Constructor-theoretic impossibility of generalising quantum protocol}

A candidate constructor for enabling the approximate transformation of a mixed state to a pure state is the universal quantum homogenizer \cite{ziman2002diluting}, first proposed as a unitary model for thermalisation. The homogenizer consists of a series of identical qubits, which interact one by one with a system qubit via a weak, unitary ``partial swap" operation. In the limit of a large number of interactions and weak coupling, the system qubit converges to the state of the homogenizer qubits, with the homogenizer qubits being arbitrarily close to their original states. 

It was recently shown that the task of transforming a qubit from a maximally mixed to a pure state via this protocol is constructor-theoretically impossible, meaning that the homogenizer deteriorates too rapidly with repeated use to be a constructor for the task \cite{violaris2021transforming,violaris2022irreversibility}. This has direct implications for the constructor-theoretic impossibility of performing the pre-processing step of the work extraction protocol for a universal input. If the input is an arbitrary, unknown pure state, then it is equivalent to a maximally mixed state, and cannot be reliably mapped to the pure $\ket{1}\bra{1}$ state via the quantum homogenizer. Similarly, if the input is an incoherent state other than $\ket{0}$ and $\ket{1}$ (which is always the case in the context of distinct coherent and incoherent states that have the same energy), then it is mixed, and again is constructor-theoretically impossible to map to $\ket{1}\bra{1}$ using the quantum homogenizer. Hence, the coherent work extraction protocol in \cite{korzekwa2016extraction} is impossible to generalise to a universal machine, even given access to powerful quantum homogenization machines. We explain the protocol and these results in more detail in Appendix \ref{homogenizer}. 

\section{Possibility of extracting identical work from non-orthogonal inputs}

What if the same amount of work is extracted from both input states? Then, the possibility of deterministic work extraction is not ruled out by the constructor-theoretic theorem. Let's imagine that the work $w_1$ that can be extracted from $\rho_1$ is less than the work $w_2$ that can be extracted from $\rho_2$, where $\rho_1$ and $\rho_2$ are arbitrary states that may or may not be perfectly distinguishable (i.e. orthogonal). Then there may exist a constructor that is able to extract the same amount of work, $w_1$, from either of the two states, without prior knowledge of the states, i.e.: $\mathbf{T_w^{\checkmark}}([\rho_1, w_1], [\rho_2, w_1])$ for arbitrary $\rho_1$ and $\rho_2$, given that $\mathbf{T_w^\checkmark}( [\rho_1, w_1])$, $\mathbf{T_w^\checkmark}( [\rho_2, w_2])$, and $w_1 \leq w_2$. More generally, the task:
\begin{equation}
\bigcup_{i}\mathbf{T_w}( [\rho_i, w_j]) 
\end{equation}
could be possible if:

\begin{equation}
\forall i: \mathbf{T_w^{\checkmark}}( [\rho_i, w_i]) \textrm{  and  } w{_i} \geq w{_j}
\end{equation}

This result does not mean that even this minimum amount of identical work is necessarily possible to extract from all the states. Let's denote the smallest amount of optimal work that can be extracted from any individual state as $w_\textrm{smallest}$. A constructor for extracting $w_\textrm{smallest}$ from that state may not necessarily be able to extract $w_\textrm{smallest}$ from the other possible input states; the result does not forbid such a protocol existing, but neither does it require it to exist. It could be interesting to demonstrate these results explicitly with a quantum protocol in future work.

\section{Further information-theoretic foundations of quantum thermodynamics}

We now discuss how constructor-theoretic work extraction and thermodynamics could extend to other settings and connect with results from resource theories. 

\subsection{Irreversibility of non-classical computation}

We conjecture that the constraint on deterministic work extraction only being possible for distinguishable inputs demonstrates, and in some respects generalises, the ``irreversibility" of quantum computation. Here the term irreversibility is used differently from the constructor-based irreversibility referred to earlier in the paper, hence we will refer to the use of irreversibility here from now on as entropic-irreversibility. 

It was shown in \cite{bedingham2016thermodynamic} that general quantum operations have an additional irreducible entropy cost to the environment as compared to classical operations. In quantum mechanics, a general quantum state $\rho$ can be considered as a distribution of states. Extending the classical definition of Shannon entropy to the quantum case, we get the von Neumann entropy $S(\rho) = - \text{Tr}(\rho \log \rho)$. Then the entropy cost of the quantum operation $\rho \rightarrow \rho '$ is based on the difference in von Neumann entropy \cite{schumacher1995quantum}, $\Delta E \geq \textrm{kTln} 2 [S(\rho) - S(\rho ')]$.

This constraint is the quantum generalisation of Landauer's principle. However, when the input quantum system can be one of at least two non-orthogonal states, and each input has a unique output, there is in general an additional entropy dissipation to the environment. A computation which saturates the Landauer bound is ``reversible", in the sense that there is no limit to how closely a complete cycle could be approached where the overall entropy change is zero (i.e. the entropy gain in the environment is perfectly balanced by the entropy change of the system). By contrast, a process that has a minimum entropy change exceeding the Landauer bound is fundamentally ``irreversible", meaning that a complete cycle cannot approach zero entropy change. The additional irreducible entropy cost to the environment for quantum operations therefore makes quantum computation entropically-irreversible.  

Similarly, theorem \ref{theorem:distinguish} from \cite{marletto2022information} shows that there are unavoidable changes in the environment for the specific task of extracting different amounts of work from non-orthogonal states. However the theorem is dynamics-independent. Understanding the possible connections between these constraints on non-classical information processing could give a dynamics-independent generalisation of the entropic-irreversibility of quantum computation to a more general form of non-classical computation. Further work in this direction could therefore lead to information-theoretic foundations of entropic-irreversibility in quantum computing, and computing with the future potential successor theories of quantum mechanics. 

\subsection{General conserved quantities}

Theorem \ref{theorem:distinguish} was derived for the simplifying assumption of a single conservation law, that of energy conservation, as is typical in thermodynamics. Development of resource theories of coherence \cite{winter2016operational,baumgratz2014quantifying}, purity \cite{horodecki2013quantumness} and entanglement \cite{chitambar2019quantum} has shown parallels between systems conserving energy and systems with other conserved quantities. There are now generalised formulations of 1$^{\textrm{st}}$ and 2$^{\textrm{nd}}$ laws of thermodynamics for systems with one or more conserved quantities, and it has been argued that energy has no special status for these laws in this respect \cite{guryanova2016thermodynamics, sparaciari2020first, lostaglio2017thermodynamic, yunger2016microcanonical}. 

Interestingly, the constructor-theoretic thermodynamics formulation lends itself to being extended for additional conserved quantities. When a quantity other than energy is conserved, the associated work medium could be defined in exactly the same way, but with its attributes having different values of the conserved quantity, rather than different values of energy. This generalised form of a work medium could then be used as a basis for formulating constructor-theoretic 2$^{\textrm{nd}}$ laws in terms of adiabatic possibility, and a theorem regarding deterministic work extraction relying on distinguishable input attributes, in precisely the same way as it was for energy. This would create a natural connection between the results regarding thermodynamics for different conserved quantities in quantum theory, and the constructor-theoretic formulation of thermodynamics. \\

We could further consider how constructor-theoretic work media are defined under multiple conserved quantities. The thermodynamic properties of a work medium should not be dependent on which one or combination of conserved quantities are used to define work. Hence, the work medium definition could be extended to account for multiple quantities being conserved simultaneously. This could give rise to a substrate-independent formulation of emergent entropy that neatly unifies entropy changes in different quantities under one 2$^{\textrm{nd}}$ law. Furthermore, since a constructor-theoretic 2$^{\textrm{nd}}$ law (based on adiabatic possibility) relies on a conserved quantity being used to define a work medium, we could explore the converse property in quantum theory: whether the 2$^{\textrm{nd}}$ laws found for different quantum resources in quantum resource theories each correspond to some kind of conservation law. This could shed light on what the underlying properties of a resource are that enable a 2$^{\textrm{nd}}$ law to be formulated for it — a question which has recently gained new life when it was shown that surprisingly, no 2$^{\textrm{nd}}$ law can exist for entanglement, though it does exist for a variety of other quantum resources \cite{lami2023no}.

\section{Conclusions and future research}

There are various approaches to the field of quantum thermodynamics, each giving insights into thermodynamics of systems in different regimes. Here we considered a case-study where there is an intersection in results from axiomatic thermodynamics, resource theories and constructor theory. We first gave a general review of these different approaches to thermodynamics, comparing some of their key features and domains of applicability. Then we analysed the problem of deterministically extracting work from coherence. We considered how the phenomenon of work being locked in coherence in a single-shot setting (work-locking), while being accessible using tomographic methods, relates to constructor-theoretic constraints from dynamics-independent laws about distinguishability. 

Building on these insights, we showed that an existing quantum protocol for deterministically extracting work from coherent states cannot be generalised to extract different amounts of work from different input states, if those inputs are non-orthogonal. This demonstrates the impossibility of a universal deterministic work extractor from coherence. We also explained how this result is implied by a recently found dynamics-independent, scale-independent theorem from constructor theory, which is based on the 1$^{\textrm{st}}$ law of thermodynamics and information-theoretic foundations. 

Additionally, we used a recently proposed quantum model for constructor-theoretically possible tasks to explicitly show the constructor-theoretic impossibility of a universal work extractor from coherence with a particular implementation, based on the quantum homogenizer.  We also considered the converse of the constructor-theoretic theorem, whereby deterministically extracting the same amount of work from non-perfectly distinguishable states can be possible. Finally we discussed potential extensions to our work, including a constructor-theoretic analysis of the entropic-irreversiblity of quantum computation, and expanding the formulation of the second law based on adiabatic possibility to multiple conserved quantities. 

In general, the usefulness of resources such as coherence in quantum thermodynamics has many subtleties and a high sensitivity to the particular assumptions and regimes being considered. We hope that demonstrating intersecting conclusions will help unify and generalise different approaches and clarify the fundamental nature and domain of applicability of results in the field. 

\section*{Acknowledgements}

We thank Chiara Marletto for discussions and comments on earlier versions of this manuscript, and Joe Dunlop for discussions about quantum operations with multiple inputs. MV thanks the Heilbronn Institute for Mathematical Research for their support.

\printbibliography

\appendix

\section{Comparing resource- and constructor-theoretic approaches} \label{resource_comparison_appendix}

\textbf{Resource theories:} Resource theories provide a general approach to formalizing scientific fields in terms of available physical states and processes \cite{coecke2016mathematical}. Physical states that are assumed to be available in unbounded number are termed \textit{free states}, because they are free to create, in the sense that no finite resources need to be used up to create them. A resource theory has a defined set of free states, and \textit{free transformations} on this set, which are transformations that only use up free states. Resource theories can be formed for any context where resources are generally classed as either finite and valuable, or unlimited and therefore free. 

Resource theories have been applied to a variety of contexts within and outside of quantum information theory, and a general framework for resource theories has been developed based on category theory \cite{coecke2016mathematical}. The resource theory of quantum thermodynamics has led to a greater understanding of nanoscale thermodynamics \cite{horodecki2013fundamental,ng2015limits}, deriving a family of free energies governing transformations \cite{brandao2013resource,brandao2015second} and limits of work extraction in the microscopic regime \cite{aaberg2013truly,skrzypczyk2014work,renes2014work}. \\

\textbf{Constructor theory:} Constructor theory provides a general, axiomatic approach to the thermodynamics of classical and quantum systems, and systems whose dynamics may be partially unknown \cite{deutsch2013constructor,deutsch2015constructor,marletto2016constructor}. While resource theories take existing dynamical laws of physics as a starting point and derive constraints on what can be done under those laws, constructor theory has its own laws of physics, e.g. the interoperability of information and the principle of locality. These laws are conjectured to be more fundamental than dynamical laws.

The central conjecture of constructor theory is that the fundamental laws of physics can be expressed in terms of principles about what tasks are \textit{possible} and \textit{impossible}, and why. In constructor theory, a task being possible means that it can be performed to arbitrary accuracy, an arbitrary number of times, and hence can be performed in a cycle. If there is some limit imposed by the laws of physics to how well a task can be performed in a cycle, then it is impossible. For example, all the particles in an isolated room moving to one corner is a dynamically allowed trajectory, but it is an impossible task, because there is no reliable way to cause the transformation in a cycle (without causing an irreducible change to the environment). 

These principles about what is possible and impossible are all exact, without dependence on probabilities or approximations. They are also scale-independent, meaning that they do not refer to a particular microscopic or macroscopic regime; and dynamics-independent, meaning they can be expressed without committing to a particular set of dynamical laws, such as those of quantum theory. The dynamics-independent properties of the principles have proved useful for devising tests for non-classical features of systems with unknown or intractable dynamics \cite{marletto2020witnessing, di2022temporal}. Meanwhile the scale-independent form of the principles means they can be used to formulate an exact form of irreversibility, whereby a task is possible and the opposite task is impossible \cite{marletto2016constructor}. It has recently been shown that this form of irreversibility is compatible with reversible dynamical laws, and demonstrated explicitly for quantum theory using a qubit model \cite{violaris2022irreversibility, marletto2022emergence}. \\

\textbf{Comparison of key features:}\\

1) State transformations vs possible tasks\\

The possibility of a task in constructor theory depends on a task being implemented in a cycle, with the machine transforming the system in a cycle called the constructor \cite{deutsch2015constructor}. The constructor should be approximately unchanged after causing the desired change in the system, and there is no limit to how well a perfect constructor for the task can be approximated. In resource theories, we can consider a catalyst for performing a transformation. 

There are various ways of defining the conditions for a catalyst in resource theories, for instance regarding the potential for correlations between the system and catalyst or arbitrarily small errors in the catalyst's return to its original state (e.g. \cite{rubboli2022fundamental}). In constructor theory, the key feature of constructors is that they retain the ability to perform a task again, rather than necessarily returning to or close to the same original state. Hence constructors are naturally formulated in terms of mappings between subspaces that enable a task, rather than retaining the same or approximately the same state. 

An interesting open problem is to extend and define the notion of catalysts used in quantum resource theories to connect with the physical laws captured by constructors in constructor theory. Recent work has drawn connections between constructor theory and the formalism of process theories, which could help towards this goal \cite{gogioso2023constructor}.\\

2) Dynamics-independence  \\

Results formulated in quantum resource theories are specifically within the context of quantum dynamics. By contrast, constructor theory has more general laws, formulated in terms of principles about possible and impossible transformations on sets. These principles are used to define a unified and generalised form of classical and quantum information, and physical laws regarding information. Building on the constructor theory of information \cite{deutsch2015constructor}, the constructor theory of thermodynamics introduces further constraints of possible and impossible tasks to define entities such as work media and processes such as work extraction \cite{marletto2016constructor}, again without committing to specific dynamics. 

Recently, the implications of constructor-theoretic principles of thermodynamics have been explored within the specific dynamics of quantum theory \cite{violaris2022irreversibility, marletto2022emergence}. In this setting, existing results and tools from quantum resource theories may have interesting connections with the constructor-theoretic possibility of thermodynamic tasks, after accounting for the subtleties regarding the above mentioned characterisation of constructor-theoretic possibility and the generalisation of catalysts to constructors.  \\

3) Connection to Lieb and Yvnangson axiomatic thermodynamics \\

It is enlightening to consider both approaches in terms of their connection to the axiomatic classical thermodynamics formulated by Lieb and Yvnangson \cite{lieb1999physics}, which itself is based on and extends works by Giles and Carathéodory \cite{giles, caratheodory1907variabilitatsbereich}. 

Resource theories are more general than Lieb and Yvnangon's axiomatic formulation of thermodynamics. They are capable of describing a family of theories that do not obey those axioms. It has been shown that a resource theory can be formulated for quantum states, precisely obeying the axioms of Lieb and Yvnangson, which turns out to be the same as the resource theory of ``noisy operations" \cite{weilenmann2016axiomatic}. Using this framework, the von Neumann entropy can be derived as the necessary quantity for ordering quantum states under these operations, forming the basis of a ``2$^{\textrm{nd}}$ law". The ordering relation between states is then given by the mathematical condition of majorization, unifying the resource theory and axiomatic approaches. 

Lieb and Yvnangson's approach is formulated around the property of ``adiabatic accessibility": the key relation between two equilibrium states is whether or not one is adiabatically accessible from the other, where adiabatic accessibility is defined operationally using a set of axioms. The overall idea is that a state is adiabatically accessible from another state if it is possible for the transformation to be implemented by some auxiliary system that returns to its original state, with the only side-effect being the lifting or lowering of a weight in a gravitational field. 

In the constructor theory of thermodynamics, the property of ``adiabatic accessibility" is reformulated as ``adiabatic possibility", where adiabatic possibility is defined solely in terms of possible and impossible tasks on a generalised form of states. The motivation of this generalisation is to achieve a scale-independent and dynamics-independent formulation of thermodynamics: adiabatic accessibility is defined in reference to a weight in a gravitational field, to generally represent any kind of work storage system. However, on microscopic and quantum scales, there are a variety of specific models for work storage systems depending on the context. Adiabatic possibility is not defined with reference to a weight in a gravitational field, but instead with reference to a ``work medium", where a work medium is itself fully defined in terms of possible and impossible tasks. This makes adiabatic possibility a scale-independent and dynamics-independent property, avoiding ambiguity regarding whether or not some specific system counts as a work storage medium. It also enable us to reason about ``weights" and work storage media that may not obey the dynamical laws of classical or quantum mechanics. 

Table \ref{table:thermo_relations} gives a summary of the approaches to ordering states according to the 2$^{\textrm{nd}}$ law in thermodynamics. This implies the following relation between the ordering of states in resource theory and constructor theory: the constructor theory of thermodynamics may coincide with a constructor-theoretic generalisation of the resource theory of noisy operations. Then the constructor-theoretic generalisation of ``majorization" could be adiabatic possibility.\\

A constructor-theoretic generalisation of the resource theory of noisy quantum operations could give rise to a generalised analog of von Neumann entropy. However generalising aspects of quantum theory to constructor theory is beyond the scope of this work. This work explores implications in the opposite direction: we take a general information-theoretic constraint on universal work extractors, proved with constructor theory, and show how it manifests in a specific protocol within quantum theory. 

\section{Distinguishability in constructor theory} \label{distinguishability}

A key aspect of information formulated using constructor theory is the constructor-theoretic definition of distinguishability \cite{deutsch2015constructor}. The physical meaning of distinguishability as defined in conventional information theory is circular: two states are distinguishable if they cause some measurement device to end up in a distinguishable pair of states, conditional on the state that was measured. So, the distinguishability of states that a system can be in is defined in terms of the distinguishability of states of the measurement device. The constructor theory of information solves this problem, giving a condition for distinguishability that does not itself refer to distinguishability, and instead only refers to the possible and impossible tasks on substrates that can instantiate information (information media).

\section{Resource-theory work extraction assumptions and results} \label{work_protocol}

The proposal in \cite{korzekwa2016extraction} has three central assumptions based on resource theories of quantum thermodynamics:

\noindent 1) The allowed transformations of the system are energy-preserving unitary operations, i.e. that commute with the total free Hamiltonian. 

\noindent 2) The average work extractable from a system $\rho_S$, incoherent in the energy eigenbasis, with Hamiltonian $H_S$ connected to a heat bath at temperature $T$, is given by the free energy difference:
\begin{equation}
    \langle W \rangle (\rho_S) = \Delta F(\rho_S) = F (\rho_S) - F(\gamma_S)
\end{equation}
where, $F(\psi) = Tr(\psi H_S) - kTS(\psi)$, $S(\psi)$ is the von Neumann entropy and $\gamma_S = e^{-H_S/kT}/Z_S$ is the thermal Gibbs state with partition function $Z_S$.

\noindent 3) Single-shot work extractable from a state $\rho_s$ is defined as,
\begin{equation}
    W_{ss}^{\epsilon}(\rho_S) = \Delta F_{0}^\epsilon (\rho_S) = F_{0}^\epsilon (\rho_S) - F(\gamma_S)
\end{equation}
where $\epsilon$ denotes a small failure probability and $\rho_S$ is again incoherent in the energy eigenbasis. 

In this framework, any work extraction protocol can be represented by a transformation of the form $\rho_S \otimes \rho_B \rightarrow \gamma_S \otimes \rho_B^\prime$, where $\rho_B$ and $\rho_B^\prime$ are initial and final incoherent battery states respectively. This is a thermal operation. It is useful to consider the effect of a dephasing channel $\mathcal{D}$ \cite{brandao2013resource}, which removes the coherence of a system and commutes with thermal operations: 

\begin{equation}
    \mathcal{D}(\psi) = \sum_i Tr(\Pi_i \psi)\Pi_i,
\end{equation}
where $\Pi_i$ are projectors onto the energy eigenspace. Then applying the three assumptions leads to the following conclusion (see \cite{korzekwa2016extraction} for the proof and more context): when there is no additional source of coherence, the maximum work extractable from a coherent state is equal to that of the decoherent state of the same energy, despite the former having more free energy. Specifically:

\begin{equation}
 \qquad \langle W \rangle (\rho_S) \leq \langle W \rangle (\mathcal{D}(\rho_S)).
\end{equation}
and similarly for single-shot work: 

\begin{equation}
    W_{ss}^{\epsilon}(\rho_S) \leq W_{ss}^{\epsilon}(\mathcal{D}(\rho_S))
\end{equation}

In this sense, there is some work ``locked" in the coherent state, that cannot be accessed without coherent resources. The question considered in \cite{korzekwa2016extraction} is then, how far can all this locked-up free energy be accessed by introducing a coherent resource?

It is shown that this ``locked work" can be accessed by introducing an ancilla system, which acts as a thermal machine. This is termed the ``reference system" $\rho_R$. It has the ladder system Hamiltonian $H_R = \sum_{n=0}^\infty n \ket{n}\bra{n}$, and its quality as a machine for the work extraction task is characterised by the coherence measure $ \langle \bar{\Delta} \rangle$ defined as: 

\begin{equation}
    \langle \bar{\Delta} \rangle = Tr_R (\rho_R \bar{\Delta}) = \frac{1}{2} Tr_R (\rho_R [ \Delta + \Delta^\dagger]).
\end{equation}
with $\Delta = \sum_{n=0}^\infty \ket{n+1}\bra{n}$ being the shift operator. The reference $\rho_R$ is coupled to a system state $\psi$, from which the work will be extracted, and a battery system $\rho_B$. The Hamiltonian of the overall system is $H = H_{S}\otimes H_R \otimes H_B$. 

\section{Caveats to coherent catalyst} \label{catalyst_caveats}

The ability to repeatedly use a coherent resource to allow transformations that change coherence of a system, without degradation of the resource, contrasts with the typical degradation undergone by quantum resources. Various caveats to the repeatability of coherent resources have been explored, arguing that coherence is not really being used repeatedly and these machines should not be termed as catalysts. 

In particular, it has been shown in \cite{vaccaro2018coherence} that while the reduced states of the outputs of the machine achieve the desired transformation, globally the outputs from multiple uses of the machine are in a very different state to that which they would be in as independent systems prepared in those reduced states. This is due to correlations between the output states affecting their global state, when they have interacted with the same machine. 

We note that for the purposes of our analysis here, the independence of the outputs of the machine is not a concern. We are comparing the effectiveness of the work extraction protocol for performing a task on an individual system state, and only the reduced state of the output is significant for our analysis. The same constraint, where only the reduced state of the output matters, is used in the constructor-theoretic analysis of quantum tasks e.g. in \cite{violaris2022irreversibility}. This is because the specific state transformation that we are considering the possibility of has a well-defined input subspace and output subspace, defined on a single copy of the system. The difference between the collection of correlated outputs and a product state of independent systems in the same reduced state does not affect the possibility of the task in question. The dependence on correlations would be significant for tasks relating to what can be done to a collection of systems prepared in a certain subspace, which is a different class of tasks. 

Instead, here we focus on a different caveat to using the catalytic properties of coherence as a thermal machine for extracting work: the impossibility of augmenting this proposal with the universality expected for typical thermodynamic machines. 

\section{Impossibility of generalising protocol with quantum homogenizer} \label{homogenizer}

The quantum homogenizer consists of a set of $N$ reservoir qubits, all initially prepared in the same state $\xi$, and accepts a system qubit in an initial state $\rho$ as input. Then the system qubit interacts with each of the reservoir qubits via a unitary partial swap operation, $U = \cos{\eta}\mathbb{1} + i \sin{\eta}\mathbb{S}$, where $\mathbb{S}$ is a SWAP operation and $\eta$ is a coupling strength parameter.

In the limit of a large reservoir (i.e large $N$), the system qubit converges to the state $\xi$, regardless of the initial states of both the system and reservoir qubits. Meanwhile, as the coupling strength of the partial swap is decreased, the reservoir qubits also remain within close vicinity of  $\xi$. Therefore, this machine essentially homogenizes all the qubits to be arbitrarily close to $\xi$, hence the name ``quantum homogenizer." 

The quantum homogenizer has recently been used as a toy-model for a demonstration of constructor-based irreversibility in quantum theory, and considered in the context of erasing the information in the memory of a quantum Maxwell's demon \cite{violaris2021transforming,violaris2022irreversibility}. 

Since the quantum homogenizer is a universal machine, it is a potential candidate for enabling the task of universal work extraction from coherence. For the work extraction, we require that an arbtirary initial input state $\rho$ is mapped to $\ket{1}\bra{1}$. We could consider using a quantum homogenizer of qubits initialised in the $\ket{1}\bra{1}$ state to perform this task for any input state $\rho$. 

The result of sequential partial swap interactions of the input qubit with the reservoir qubits is then: 
\begin{equation}
    U_N^\dagger \dots U_1^\dagger (\rho\otimes\ket{1}\bra{1}^{\otimes N}) U_1 \dots U_N \approx \ket{1}\bra{1}^{\otimes N+1}.
\end{equation}
where the unitary $U_k = U \otimes (\bigotimes_{j\neq k} \mathbb{1}_j)$ represents interaction of the input qubit with the k$^{\textrm{th}}$ reservoir qubit. Note that all the output states will be arbitrarily close to $\ket{1}\bra{1}$ as $\eta \rightarrow 0$, and this joint transformation does not violate linearity or even unitarity due to the slight deterioration of the reservoir qubit states \cite{ziman2002diluting}. 

Then the key question becomes: can the quantum homogenizer be re-used to reliably perform this transformation an arbitrary number of times? If yes, then this would enable a universal deterministic work extractor from coherence. 

The condition to perform a task to arbitrary accuracy an arbitrary number of times is precisely the definition of a constructor in quantum mechanics. Previous work analysing the quantum homogenizer as a constructor has demonstrated that while it functions as a constructor when transforming a qubit from a pure state to a mixed state ($T = \{ \text{\textit{pure state}} \rightarrow \text{\textit{mixed state}} \}$), it is not a constructor for the transpose task of transforming a mixed state to a pure state ($\tilde{T} = \{ \text{\textit{mixed state}} \rightarrow \text{\textit{pure state}} \}$)  \cite{violaris2022irreversibility}. Hence the former task is constructor-theoretically possible, while the latter is not necessarily possible, and is manifestly not possible using the quantum homogenizer. 

In the protocol for extracting work from coherence, we need to transform a pure state with coherence in the computational basis (eq \ref{eq:CoherenceInputstate}) arbitrarily close to $\ket{1}$, also a pure state. 

\textbf{Pure, distinct and non-orthogonal input states.} Assume that some operation can transform an arbitrary pure state, into a fixed pure state, let's say $\ket{1}$. Since the operation must work for any pure input state, without prior knowledge of which state it is, the input is equivalent to a maximally mixed state. For example, if the input could be either $\ket{0}$ or $\ket{1}$, with no prior information about which of those states it is, then we describe the input by the mixture $\rho = \frac{1}{2}(\ket{0}\bra{0}+ \ket{1}\bra{1})$. By extension, if the input could be any single-qubit pure state, with no information about which state it is, it is also described by the maximally mixed state. 

Hence, the possibility of mapping any pure state to a fixed pure state reduces to the possibility of transforming a maximally mixed state to a fixed pure state, i.e. the task $\tilde{T} = \{ \text{\textit{mixed state}} \rightarrow \text{\textit{pure state}} \}$. However, it has been shown that the quantum homogenizer is not a constructor for this task, meaning that the task is not constructor-theoretically possible using the quantum homogenizer. 

Therefore universal work extraction from coherence cannot be reliably implemented with any known constructor, and is ruled out by the conjecture that the mixed-to-pure task is constructor-theoretically impossible. A universal work extractor from coherence using this protocol is constructor-theoretically possible if and only if the task $\tilde{T} = \{ \text{\textit{mixed state}} \rightarrow \text{\textit{pure state}} \}$ is also possible. 

As the impossibility is conjectured to hold for any task transforming a mixed state to a less mixed state, the impossibility also applies for a mapping of an arbitrary pure state to a state within the vicinity of a fixed pure state (which has an equivalent description as a non-maximally mixed state). 

\textbf{Incoherent state.} When a coherent and incoherent state are distinct and have the same energy, the incoherent state is mixed (i.e. when the incoherent state is not $\ket{0}$ or $\ket{1}$). Therefore, using the quantum homogenizer to map the state to within the vicinity of the pure state $\ket{1}$ is also a constructor-theoretically impossible task. Hence, the coherent work extraction protocol in\cite{korzekwa2016extraction} is impossible to generalise to a universal machine even when given access to powerful quantum homogenization machines. 

\end{multicols}
\end{document}